\def\re#1{(\ref{#1})}
\def\beq{\begin{equation}}
\def\eeq{\end{equation}}
\def\beeq{\begin{eqnarray}}
\def\beeqn{\begin{eqnarray*}}
\def\eeeq{\end{eqnarray}}
\def\eeeqn{\end{eqnarray*}}

                  \def\G{\Gamma}
\def\de{\delta}

\def\l{\lambda}                 
\def\m{\mu}
\def\n{\nu}
\def\ta{\tau}

\def\s{\sigma}                  
\def\th{\theta}

\def\z{\zeta} 

\newcommand{\KK}{{\cal K}}

\newcommand{\OO}{{\cal O}}
\newcommand{\PP}{{\cal P}}

\newcommand{\lp}{\left(}
\newcommand{\rp}{\right)}
\renewcommand{\lq}{\left[}
\renewcommand{\rq}{\right]}

\newcommand{\no}{\nonumber}
\newcommand{\ph}{\phantom} 

\def\tr{\,\mbox{Tr}\,}
\def\frac#1#2{ {{#1} \over {#2} }}

\def\half{\mbox{\small $\frac{1}{2}$}}
\def\p{\partial}

\def\eg{\hbox{\it e.g.}{ }}

\documentclass[letterpaper,notoc,12pt]{JHEP3} 

\usepackage{epsfig,multicol}
\newcommand\fverb{\setbox\pippobox=\hbox\bgroup\verb}
\newcommand\fverbdo{\egroup\medskip\noindent%
			\fbox{\unhbox\pippobox}\ }
\newcommand\fverbit{\egroup\item[\fbox{\unhbox\pippobox}]}
\newbox\pippobox

\title{Two-dimensional non-commutative Yang-Mills theory: coherent
effects in open Wilson line correlators}
\author{A. Bassetto\thanks{Partially supported by the European Community 
network HPRN-CT-2000-00149.}~ , G. De Pol and
F. Vian\\
Dipartimento di Fisica ``G.Galilei", Via Marzolo 8, 35131
Padova, Italy\\
INFN, Sezione di Padova, Italy\\
E-mail: \email{bassetto@pd.infn.it},\email{depol@pd.infn.it}, \email{vian@pd.infn.it}}
\received{} 		
\revised{}
\accepted{}		

\preprint{DFPD 03/TH 19}	

\abstract{
A perturbative calculation of the correlator of three parallel open Wilson
lines is performed for the $U(N)$ theory in two non-commutative 
space-time dimensions. In
the large-$N$ planar limit, the perturbative series is fully resummed and asymptotically leads 
 to an exponential increase of the correlator with the
lengths of the lines, in spite of an  interference effect between
lines with the same orientation. This result generalizes a similar
increase occurring in the two-line correlator and is likely to persist
when more lines are considered provided they share the same direction.
}

	
\dedicated{}

\begin{document} 

\section{Introduction}

Interesting dynamical information on gauge theories defined on noncommutative
space-time are provided by correlation functions of gauge invariant operators
\cite{observ}.

Noncommutativity of $D$-dimensional Minkowski 
space-time is encoded in a real antisymmetric matrix 
$\theta^{\mu\nu}$:
\beq
\label{alge}
[x^\mu,x^\nu]=i\theta^{\mu\nu}\quad\quad\quad\quad \mu,\nu=0,..,D-1
\eeq
and a $\star$-product of two fields $\phi_1(x)$ and $\phi_2(x)$  can be defined
by means of Weyl symbols
\beq
\label{star}
\phi_1\star\phi_2(x)=\int \frac{d^Dp\, d^Dq}{(2\pi)^{2D}}\exp \lq -\frac{i}2 \, 
p_{\mu}\theta^{\mu\nu}q_{\nu}\rq \exp(ipx) \tilde \phi_1(p-q) \tilde \phi_2(q).
\eeq
Then noncommutative gauge theories (NCGT)~\footnote{Reviews on NCGT may be found for instance in \cite{rev}.} are most easily formulated by replacing the 
usual multiplication of fields in the Lagrangian with the Moyal
$\star$-product. 
The resulting action
makes them obviously non-local. 
As a consequence, gauge invariance in this case (star-gauge invariance) 
entails an integration over space-time variables and the
possibility of having {\it local} probes is lost.  

There is however a remarkable recipe in NCGT which turns
local operators into gauge invariant observables carrying a non-vanishing
momentum \cite{IIKK}. {\it Open} Wilson lines $W(p)$
with momentum $p_{\mu}$ can be considered which are
gauge invariant provided their length $l^{\nu}$
is related to the momentum as follows
\beq\label{length}
l^{\nu}=p_{\mu}\theta^{\mu\nu}.
\eeq

Then, averaging any usual {\it local} gauge invariant operator with respect to space-time
and group variables with a weight given by an open Wilson line  provides indeed 
an (over-complete) set of dynamical observables.
The Wilson line itself
provides the simplest example of such an average, the {\it local} operator  
being just unity in this case \cite{IIKK,gross}.

In \cite{gross} a perturbative estimate in the 't Hooft limit of open Wilson line
correlators was carried out in four dimensions for a ${\cal N}=4$ supersymmetric 
$U(N)$ NCGT, with $\theta$ of a ``magnetic'' type.
In four dimensions ladder diagrams dominate. 
An all order resummation was compared with a dual 
supergravity result and a good agreement was found.
The correlator of {\it two} (necessarily parallel) lines did exhibit an 
exponential increase
with the length of the lines   
whereas the {\it normalized} correlators of multiple Wilson lines
were found to be exponentially decreasing. This result was ascribed to the
lack of parallelism (and thereby of coherence) which generally occurs 
when several lines are considered.

In two dimensions noncommutativity necessarily involves
the time variable, but the Lorentz symmetry is not violated owing to the
tensorial character of $\theta^{\mu\nu}$.  
In the light-cone gauge, the perturbative calculation is greatly 
simplified, thanks to the decoupling of Faddeev-Popov ghosts
and to the vanishing of the vector vertices. 
When considering the correlator of two open Wilson lines, 
it turns out that in all diagrams  which are leading in the large-$N$ limit, 
the  contributions of $\th$-dependent phases cancel after volume integration \cite{noi}.
This is far from trivial because cyclic permutations 
inside the trace associated to each Wilson line do not entail any change in the
related colour factors; as a consequence leading diagrams as far as
colour is concerned are not necessarily of a ladder type. Nevertheless
it turns out they contribute in the limit of large length of the lines
on an equal footing with
the simpler ladder graphs; this is the reason why such diagrams were
still called  ``planar'' in \cite{noi}.
There a detailed perturbative calculation of the correlator of two
Wilson lines was performed in two dimensions    
in the 't Hooft limit $N\to \infty$, $g^2N$ fixed.
After a full
resummation of the perturbative series, an exponential increase of the
correlator with respect to the length of the lines was found, in
agreement with the analogous behaviour occurring in four dimensions.
Such a result was also confirmed by a non perturbative calculation on
a bidimensional torus using Morita equivalence, followed by
decompaction, in a suitable region of the parameters involved,
corresponding to the planar regime as defined in \cite{Mandal}.

We stress that in order to reach this goal, the large-$N$ limit was essential.
Indeed, in the perturbative calculation a plethora of diagrams,
subleading as far as colour is concerned, was disregarded; on the
non-perturbative side, a particular saddle-point approximation, describing
the mentioned planar phase of the theory, was crucial. Such
an approximation requires the limit $N\to \infty$, with $N$ larger
than the winding number associated to the momentum carried by the lines. As a consequence the conclusions
in \cite{pania} do not apply to the treatment in \cite{noi}. In
passing we notice that 
an analogous regime was considered in the four-dimensional case \cite{gross}.
 
It is the purpose of this paper to generalize the perturbative
calculation in \cite{noi}
to correlators of three {\it parallel} Wilson lines in the same ``planar''
context. This is not an academic
task, since multiple line correlators in a generic configuration 
are not expected to increase with
the length of the lines on the basis of the estimate in \cite{gross}.
Remarkably, we find instead they keep increasing, when the lines
are parallel, at the same rate as
the two-line correlator, in spite of the fact
that exchanges between  lines with the same orientation
generate an interference effect. This is indeed the case,
but is overwhelmed by the coherent increase
due to parallelism of lines with opposite orientation.

As a consequence, if a decrease would occur, it should be ascribed only
to the lack of parallelism; unfortunately, we are unable to extend our calculation
to this case and therefore  have no new prediction concerning
non-parallel lines. 

The outline of the paper is as follows.
In section 2 we define the three-line correlator and introduce the basic
quantities and notations 
we will use throughout the paper. Section 3 is concerned with the perturbative
calculation at any order of a generic configuration of the diagrams with
the only restriction to {\it parallel} Wilson lines. In section 4 the
perturbative 
series are concretely summed obtaining an asymptotic expression for the
increase of the correlator with the line lengths. Final considerations
and comments concerning possible future developments are considered in the
conclusions, while technical details are deferred to the appendices A and B.

\section{The three-line correlator}

The classical action of the $U(N)$ Yang-Mills theory on a noncommutative
two-dimensional 
space is
\beq 
\label{act}
S=-\frac12 \int d^2x\, \tr F_{\m\n} \star  F^{\m\n}\,,
\eeq
where the field strength $F_{\m\n}$ is given by
\beq
F_{\m\n}=\p_\m A_\n -\p_\n A_\m -ig (A_\m\star A_\n - A_\n\star A_\m)
\eeq
and $A_\m=A_\m^a T^a$ is a $N\times N$ matrix, with $T^a$ normalized
as follows: $\tr T^a T^b=\half \de^{ab}$, $a,b$ denoting $U(N)$ indices.

The action    eq.~\re{act} is invariant under infinitesimal $U(N)$
noncommutative gauge transformations 
\beq
\label{gauge}
\de_\l A_\m= \p_\m \l -i g (A_\m\star\l -\l\star A_\m) \,.
\eeq
As noticed in \cite{gross}, under this transformation 
the operator $\tr F^2(x)$ is not left invariant 
\beq\label{opera}
\tr F^2(x) \longrightarrow \tr U(x) \star  F^2(x) \star U^\dagger (x)\,,
\eeq
with $U(x)=\exp_* (ig \l (x))$. To recover a gauge invariant operator,
one has to integrate over all space, since  
$\star$-products  inside integrals can be cyclically permuted.

A Wilson line of length $l$ can be  defined by
means of the Moyal product as \cite{observ}  
\beq
\label{wline}
\Omega_\star[x,C]=P_{\star} \exp \lp
ig\int_0^l A_\m 
(x+\z(\s))\, d\z^\m(\s)\rp \,,
\eeq
where $C$ is the curve 
parameterized by $\z(\s)$, with $0 \leq \s \leq 1$, $\z(0)=0$,
$\z(1)=l$, and $P_\star$
denotes noncommutative path ordering along $\z(\s)$ from right to left
with respect to increasing $\s$ of $\star$-products of functions.
The Wilson line is not invariant under a gauge transformation
\beq\label{linetrans}
\Omega_\star [x, C] \longrightarrow  U(x) \star \Omega_\star [x,C] \star 
U^\dagger (x+l)\,.
\eeq
In order to recover gauge invariance one should perform a complete trace
operation, which, in a noncommutative context, entails also integration over
coordinates.

The following operator
\beq \label{linedef0}
W(p, C)= \int d^2x \, \tr \Omega_\star[x, C]\star e^{ipx}\,,
\eeq
turns out to be invariant provided $C$ satisfies the condition
\beq\label{endpoints}
l^\n=p_\m \th^{\m\n}
\eeq
(the Wilson line extends in the direction transverse to the momentum).
The particular case $p_{\mu}=0$ corresponds to a closed loop.

For simplicity in the following only straight lines will be considered.
Then one can easily realize that any local operator $\OO (x)$ in ordinary
gauge theories admits a noncommutative generalization
\beq \label{ope}
{\tilde \OO (p)}= 
\tr \int d^2x \, \OO (x) \star \Omega_\star [x, C] 
\star e^{ipx}\,,
\eeq
each of the ${\tilde \OO (p)}$'s being a genuinely different operator
at different momentum.

Remarkably, owing to eq.~\re{endpoints}, at large values of $|p|$, 
gauge invariance requires that
the length of the Wilson line  becomes large. This fact can be interpreted
as a manifestation of the UV-IR mixing phenomenon.

The authors of \cite{noi} studied the two-point function
$\langle W(p,C)W^\dagger (p,C')\rangle$ in two space-time dimensions, 
where $W(p,C)$ has been
defined via eqs.~\re{wline} and \re{linedef0}. It represents the
correlation function of two straight (anti-)parallel Wilson lines of equal
length, each carrying a transverse momentum $p$.

The theory defined through eq.~\re{act} was quantized in the
light-cone gauge $A_-=0$ at {\em equal 
times}, the free propagator having the following causal expression
(WML prescription)
\begin{equation}
\label{WMLprop}
D^{WML}_{++}(x)={1\over {2\pi}}\,\frac{x^{-}}{-x^{+}+i\epsilon x^{-}}\,,
\end{equation}
first proposed by T.T. Wu \cite{Wu}. In turn this propagator is nothing
but the restriction in two dimensions of the expression proposed 
by S. Mandelstam and G. Leibbrandt \cite{Leib} in four
dimensions~\footnote{In dimensions higher than two, where physical
degrees of freedom are switched on (transverse ``gluons''), this
causal prescription is mandatory \cite{libro}.}  and 
derived by means of a canonical quantization in \cite{bosco}.

This form of the propagator allows a smooth transition to an Euclidean 
formulation: $x_0\to ix_2, x_1\to x_1$. The propagator above becomes
\begin{equation}
\label{euprop}
D_{E}(x)={1\over {2\pi}}\,\frac{x_1+ix_2}{x_1-ix_2}= -\frac{1}{2\pi^2}
\int d^2 k e^{ikx}\frac{1}{(k_1-ik_2)^2}\,.
\end{equation}
In the sequel momentum integrals will always be performed by means of a
``symmetric integration'' \cite{Wu}, namely by an angular average around
the pole. We shall also use the light-cone notation $k_-\equiv -(k_1-ik_2)$.

The light-cone gauge gives rise to
other important features like the decoupling of Faddeev-Popov ghosts,
which occurs also in the noncommutative case \cite{Das}, and
the absence  of the triple gluon vertex in
two dimensions.
Consequently the computation of the Wilson line correlator
is enormously simplified.

The main result   found in \cite{noi} is the exponential increase of the two-line correlator $<W(p,C)W^{\dagger}(p,C')>$ with respect to the (equal) length 
of the lines $C$ and $C'$.

We now consider the three-line correlator 
$<W(p_1,C)W^{\dagger}(p_2,C')W(p_3,C'')>$; since we choose the lines to be
parallel, momentum conservation $$p_1+p_2+p_3=0$$ implies the relation
\beq \label{leng}
l_1+l_3=l_2
\eeq 
among the three lengths.

With no loss of generality we choose  the paths
stretching along $x^0$, so that $p_{i}$ points in the spatial direction.
The parallel paths $C'$ and $C''$ may be chosen at a distance $\Delta'$ and
$\Delta''$ from $C$, respectively. However these distances amount only 
to the appearance of two irrelevant phase factors (see \cite{noi}).
As a consequence it is not restrictive to imagine the three lines
superimposed and the dependence on $C$, $C'$ and $C''$ will be understood
in the sequel.

The variables $\z$, $\xi$ and $\eta$, parameterizing the lines $C$,
$C'$ and $C''$, respectively,   can be conveniently  rescaled as
$\z=\s l_1$, $\xi=\s' l_2$ and $\eta=\s'' l_3$, with $0\le\s, \s', \s''\le 1$
and $l_i^0=|\theta p_i|, i=1,.,3$, 
according to eq.~(\ref{length}).

We begin by expanding the line operators
\beeq \label{nuopertline}
&&W(p_1)\,
=\,
\sum_{m=0}^\infty \,(igl_1)^m \int d^2x\int_{\s_m>\ldots>\s_1}
[d\s]\, \tr
A(x+\z_1) \star  \ldots \star A(x+\z_m)\star e^{ip_1x}
\no\\
&&W^\dagger(p_2)\,
=\,
\sum_{m=0}^\infty \,(-igl_2)^m \int d^2x
\int_{\s'_m>\ldots>\s'_1} [d\s']\, \tr
A(x+\xi_m)  \star \ldots \star A(x+\xi_1)\star
e^{-ip_2x}\no\\
&&
W(p_3)\,
=\,
\sum_{m=0}^\infty \,(igl_3)^m \int d^2x\int_{\s''_m>\ldots>\s''_1}
[d\s'']\, \tr
A(x+\eta_1) \star  \ldots \star A(x+\eta_m)\star e^{ip_3x}\,.\no \\
\eeeq
The perturbative calculation now continues with performing suitable contractions
among the operators in these equations, giving rise only to products of propagators
since no internal vertices are present in such a theory. At this stage it is essential
to consider the large-$N$ limit which allows to disregard a large part of 
({\it a priori} possible) configurations; we explicitly assume that non-leading 
configurations, as far as colour is concerned, can be consistently dropped.

One should keep in mind that cyclic permutations leave traces invariant
and therefore several leading equivalent configurations are to be
taken into account. Moreover a given propagator may either connect two
different lines or start and end on the same line. This last situation will
be treated separately. Finally, disconnected diagrams will not be considered;
as a matter of fact they contribute only for particular values of the line momenta
and trivially factorize in the product of lower order correlators.
Still, connected diagrams are colour subleading with respect to disconnected
ones as it will become apparent shortly. As a consequence, when normalized
to two-line correlators, they vanish as $1/N$ in the 't Hooft limit. 
Nevertheless, if instead the coupling $g$ is kept fixed and $N$ is
sent to $\infty$, the normalized three-line correlator becomes
sizeable, as will be cleared in the sequel.

\section{The perturbative calculation}

Now we start performing contractions between operators giving rise to
propagators attached to the three open lines. 
For the time being we postpone the discussion of diagrams with propagators beginning
and ending on the same line; we will comment upon them later on.

First of all we have to single out configurations which are leading as far as colour
is concerned. We denote with $n_{ij}$ the number of propagators exchanged
between the line $i$ and the line $j$. If we introduce the shorthand notation
$$(a_1\ldots a_k)\equiv Tr(T^{a_1}\ldots T^{a_k})\,,$$
we can easily realize
that the following cyclic pattern
\begin{eqnarray}
\label{lead}
&&(a_1\ldots a_{n_{13}}b_1\ldots b_{n_{12}})(b_{n_{12}}\ldots b_1c_1\ldots c_{n_{23}})
(c_{n_{23}}\ldots c_1a_{n_{13}}\ldots a_1)=\frac{1}{N}\cdot (N/2)^{n},\no \\
&&\qquad\qquad n\equiv n_{12}+n_{23}+n_{13},
\end{eqnarray} 
is colour leading. 

In order to derive the equation above (and suitable generalizations), one
has to perform the operations of joining and splitting different strings
of colour matrices. The basic relation to be used, in the case of $U(N)$, reads
\beq\label{joi}
T^a_{ij}T^a_{kl}= \frac{1}{2}\delta_{il}\delta_{jk},
\eeq
where the index $a$ is to be summed over.
Then, for instance, one gets
\beeq\label{stri}
&&(a_1\ldots a_{n_{13}}b_1\ldots b_{n_{12}})(b_{n_{12}}\ldots b_1c_1\ldots c_{n_{23}})
(c_{n_{23}}\ldots c_1a_{n_{13}}\ldots a_1) \no \\
&&= \frac{1}{4}(a_1\ldots a_{n_{13}}b_1\ldots b_{n_{12}-1}b_{n_{12}-1}
\ldots b_1c_1\ldots c_{n_{23}-1}
c_{n_{23}-1}\ldots c_1a_{n_{13}}\ldots a_1)
\eeeq
and the result eq.~\re{lead} easily follows.

\smallskip
These cyclic configurations are to be summed  and
we have now to consider the geometrical structure they entail. 

Starting from eq.~\re{nuopertline}, we contract the $A$'s in such a 
way that the resulting diagram with $n_{12}$, $n_{13}$, $n_{23}$
propagators is of leading order in $N$, corresponding
to the colour pattern in eq.~\re{lead}. After analytic continuation to
Euclidean variables, using eq.~\re{euprop}, we get at a fixed
perturbative order $n=n_{12}+n_{13}+n_{23}$
\beeq
\label{generic}
&&I_{n_{12},n_{13},n_{23}}=(-1)^{n_{12}+n_{23}}N^{n-1} \lp
\frac{g^2}{4\pi^2}\rp^n  
\int \prod_{i=1}^{n_{12}} 
d\z_i\, d\xi_i\, 
\prod_{j=1}^{n_{23}} d\xi'_j\, d\eta_j\, 
\prod_{k=1}^{n_{13}} d\z'_k\, d\eta'_k \no \\
&&\times
\int \prod_{i=1}^{n_{12}} e^{-i P_i \cdot (\xi_i-\z_i)} 
\frac{d^2 P_i}{P_{i-}^2} \,
\prod_{j=1}^{n_{23}} e^{-i Q_j \cdot (\eta_j-\xi'_j)} \frac{d^2
Q_j}{Q_{j-}^2} \,
\prod_{k=1}^{n_{13}} e^{-i R_k \cdot (\eta'_k-\z'_k)} \frac{d^2
R_k}{R_{k-}^2}\,e^{i \PP (P_i, Q_j, R_k; \th, {\bf p}) }\no\\
&& \times \,
\de^{(2)}(\sum_i P_i +\sum_k R_k -p_1) \, \de^{(2)}(\sum_j Q_j -\sum_i P_i -p_2) \,
\de^{(2)}(\sum_k R_k  + \sum_j Q_j + p_3) \,,\no\\
&&\hfill
\eeeq
where ${\bf p}=\{p_1,p_2,p_3\}$, the integrations over the line
variables are ordered  according to the colour pattern in
eq.~\re{lead}  and the Moyal phase $\PP (P_i, Q_j, R_k;\th, {\bf p})$ 
is a linear function of $P_i$, $Q_j$ and $R_k$, depending on the
topology.

A typical example is explicitly provided in figure 1.
\begin{figure}[h]
\label{3+2+2}
\begin{center}
\epsfxsize=6cm
\epsffile{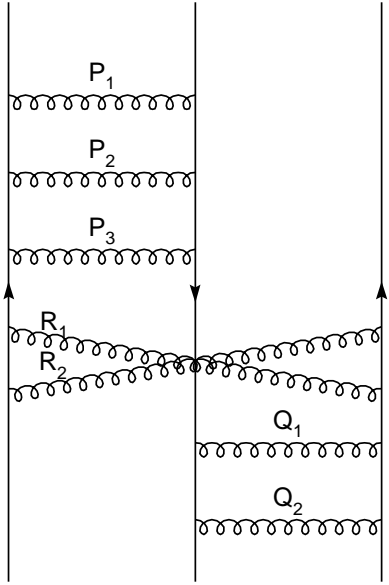}
\caption{A non-trivial leading diagram in the large-$N$ limit.}
\end{center}
\end{figure}
The corresponding analytical expression is
\beeq
\label{generico}
&&I_{3,2,2}=-N^{6} \lp
\frac{g^2}{4\pi^2}\rp^7  
\int \prod_{i=1}^{3} 
d\z_i\, d\xi_i\, 
\prod_{j=1}^{2} d\xi'_j\, d\eta_j\, 
\prod_{k=1}^{2} d\z'_k\, d\eta'_k 
\int \prod_{i=1}^{3} e^{-i P_i \cdot (\xi_i-\z_i)} 
\frac{d^2 P_i}{P_{i-}^2} \no \\
&&\times
\prod_{j=1}^{2} e^{-i Q_j \cdot (\eta_j-\xi'_j)} \frac{d^2
Q_j}{Q_{j-}^2} \,
\prod_{k=1}^{2} e^{-i R_k \cdot (\eta'_k-\z'_k)} \frac{d^2
R_k}{R_{k-}^2}\,e^{-\frac{i}2 \lq p_3 \th \lp P_1+P_2+P_3+R_1+R_2\rp \rq
}\\
&& \times \,
\de^{(2)}(\sum_i P_i +\sum_k R_k -p_1) \, \de^{(2)}(\sum_j Q_j -\sum_i P_i -p_2) \,
\de^{(2)}(\sum_k R_k  + \sum_j Q_j + p_3) \no \,.
\eeeq

Following a simple but rather lengthy procedure, which is fully
described in appendix~A, we can rewrite eq.~\re{generic} in a form in
which factorization of propagators in coordinate variables is manifest
\beeq
\label{fourier}
&&I_{n_{12},n_{13},n_{23}}
=\frac{(-1)^{n_{13}}}{(2\pi)^6 N} 
\lp \frac{g^2 N}{4\pi}\rp^n  \de^{(2)}(p_1+p_2+p_3) 
\int \prod_{i=1}^{n_{12}} 
d\z_i\, d\xi_i\, 
\prod_{j=1}^{n_{23}} d\xi'_j\, d\eta_j
\no \\
&&\ph{I_{n_{12},n_{13},n_{23}}}
\times 
\prod_{k=1}^{n_{13}} d\z'_k\, d\eta'_k 
\int d^2x\,  d^2y\,  d^2z\,  
\prod_{i=1}^{n_{12}} \frac{x-a_i}{(x-a_i)^*} 
\prod_{j=1}^{n_{23}} \frac{y-b_j}{(y-b_j)^*} 
\prod_{k=1}^{n_{13}} \frac{z-c_k}{(z-c_k)^*}\no \\
&&\ph{I_{n_{12},n_{13},n_{23}}}
\times 
\int  d^2P\,  
e^{i P \cdot (x+y+z)} e^{-\frac{i}3\lq (p_1 - p_2)\cdot x+(p_2 -
p_3)\cdot y +  (p_3 - p_1)\cdot z \rq} \,,  
\eeeq
where $a_i=\xi_i-\z_i + A_i ({\bf \th p})$, $b_j=\eta_j-\xi'_j + B_j (
{\bf \th p})$,
$c_k=\eta'_k-\z'_k + C_k ({\bf \th p})$ and $A_i$, $B_j$, $C_k$ are 
linear combinations of $\{\th p_l\}_{l=1,2,3}$, obeying $
\sum_i P_i A_i+\sum_j Q_j B_j + \sum_k R_k C_k = -
\PP (P_i, Q_j, R_k;\th, {\bf p})$.

Once we have recognized that the structures in eq.~\re{lead} 
allow factorization of propagators
in coordinate variables, a little thought is enough to conclude that cyclic
permutations on a line do not alter such a pattern. As a matter of fact
any open line with $n$ attached propagators can be considered as a diagram with 
$n+1$ vectors, the extra vector $p=\sum p_j$ representing the momentum balance. 
We know that
the Moyal phase associated to {\it
planar} diagrams  only depends on external momenta and is cyclically invariant 
thanks to total momentum conservation \cite{Minwalla}. One can wonder what happens
if only the $n$ vectors $p_j$ are cyclically permuted. Since the phase is quadratic
in the momenta and its variation must vanish at $p=0$, it is easy to conclude
that its variation is at most {\it linear} with respect to $p_j$. 
As a consequence the factorization property we mentioned is unaffected and all
the configurations obtained from cyclic permutations over each line are to be
considered on an equal footing, counted and summed as well.  
We stress that, remarkably, factorization of propagators in coordinate
variables  occurs just in those diagrams which are dominant at
large $N$. This feature in turn makes $\th$-dependence trivial, as it
will be shortly cleared, since $\th$ intervenes just through the length
of the lines. 

We find
this factorization very peculiar; it is indeed characteristic of two-dimensional
theories and
is not fulfilled, for instance, in the four-dimensional case where planarity just 
means {\it ladder}
diagrams \cite{gross}.

\smallskip
Coming back to eq.~\re{fourier},
integration over coordinates can be performed (see again
appendix~A). Symmetric integration provides the natural regularization. For instance, we have 
\beq
\label{intp}
\int d^2x\,
\prod_{i=1}^{n_{12}} \frac{x-a_i}{(x-a_i)^*} \, e^{i(P-\tilde{p}_{1})\cdot x}
=\sum_{m=1}^{n_{12}} 
\frac{\pi (2i)^{m+1} m!}{ (P-\tilde{p}_{1})_-^{m+1}} \sum_{i=1}^{n_{12}}
e^{i(P-\tilde{p}_{1})\cdot a_i} \,V(m,i,{\bf a})\,,
\eeq
where
\beq
\label{v} 
V(m,i,{\bf a})= 
\sum_{j_1<j_2<\ldots<j_{n_{12}-m}}
\frac{(a_i - a_{j_1})\ldots (a_i - a_{j_{n_{12}-m}})}{\prod_{j\neq
i}(a_i-a_j)^*} \,,
\eeq
having defined
${\bf a}=\{a_1,\ldots,a_{n_{12}}\}$, 
$\tilde{p}_{1}=\frac{p_1 -p_2}3$, and similarly, for integration over $y$ and $z$, 
$\tilde{p}_{2}=\frac{p_2 -p_3}3$,
$\tilde{p}_{3}=\frac{p_3 -p_1}3$.
Eventually, we have to perform the integration over $P$, following
appendix~A. Collecting all
the terms depending on $P$ in eq.~\re{fourier}, we
get 
\beeq
\label{p-int}
&&\int d^2P\, \frac{e^{iP\cdot (a_i+b_j+c_k)}}{ (P-\tilde{p}_{1})_-^{m_1+1}
(P-\tilde{p}_{2})_-^{m_2+1} (P-\tilde{p}_{3})_-^{m_3+1}} \equiv 
\KK (m_1,m_2,m_3;\tilde{p}_{1}, \tilde{p}_{2}, \tilde{p}_{3}; a_i+b_j+c_k)
\no
\\
&& = 4 \pi 
\sum_{l_2,\, l_3=0}^{m_1}
{m_2+l_2 \choose l_2}  \,
\frac{(-1)^{l_2} }{(\tilde{p}_1 -\tilde{p}_2 )^{m_2+l_2+1}_-}\,
{m_3+l_3 \choose l_3} \, 
\frac{(-1)^{l_3} }{(\tilde{p}_1 -\tilde{p}_3 )^{m_3+l_3+1}_-}\no \\
&& \ph{4 \pi \sum_{l_2,\, l_3=0}^{m_1}} 
\times \frac{\lp\frac{i}2 \rp^{m_1-l_2-l_3+1} }{(m_1-l_2-l_3)!} \,
\frac{(a_i+b_j+c_k)^{m_1-l_2-l_3}}{(a_i+b_j+c_k)^*} +
(1\leftrightarrow2)
+  (1\leftrightarrow3)\,.
\eeeq
Exploiting eq.~\re{intp}, we have
\beeq
\label{vvv}
&&I_{n_{12},n_{13},n_{23}}
=\frac{(-1)^{n_{13}}}{(4\pi)^3 N} 
\lp \frac{g^2 N}{4\pi}\rp^n  \de^{(2)}(p_1+p_2+p_3) 
\int \prod_{i=1}^{n_{12}} 
d\z_i\, d\xi_i\, 
\prod_{j=1}^{n_{23}} d\xi'_j\, d\eta_j\, 
\no \\
&&\times
\prod_{k=1}^{n_{13}} d\z'_k\, d\eta'_k 
\sum_{m_1=1}^{n_{12}} \sum_{m_2=1}^{n_{23}} \sum_{m_3=1}^{n_{13}} 
(2i)^{m_1+m_2+m_3+3} m_1!m_2!m_3! 
\sum_{i=1}^{n_{12}} \sum_{j=1}^{n_{23}} \sum_{k=1}^{n_{13}} 
V(m_1,i,{\bf a}) 
\no\\
&&\times  V(m_2,j,{\bf b}) \, V(m_3,k,{\bf c})\,
\KK (m_1,m_2,m_3;\tilde{p}_{1}, \tilde{p}_{2}, \tilde{p}_{3}; a_i+b_j+c_k)\,.
\eeeq
Finally, we need to carry out some algebra.
We will show in appendix~A that the only  contributions to  the
summations above are those, referring to eq.~\re{p-int},   
obeying  the conditions
$m_r+l_r+1=2$, $r=2,3$ for the first term  of the r.h.s., and
analogous ones for the other two terms. 
One can see that the terms surviving are
independent of the  variables $a_i$, $b_j$ and $c_k$. 
It follows that integration over geometric variables in eq.~\re{vvv}
is trivial. Just to fix ideas, integration over 
$\z_i$, $\z'_k$,
produces a factor  $l_1^{n_{12}+n_{13}}/(n_{12}+n_{13})!$, and so on. 
Thus, we end up with the following
expression for eq.~\re{vvv}  ~\footnote{The momentum dependence
looks particularly simple thanks to our choice of coordinates in which
momenta are taken along the $x$-direction. A more general choice would
entail a phase in intermediate steps which however cancels in the final result
(see \cite{noi}).} 
\beeq
\label{fourier2}
&&I_{n_{12},n_{13},n_{23}}
=\frac1{\pi^2 N} \, \de^{(2)}(p_1+p_2+p_3) 
\lp \frac{g^2 N l_1 l_2}{4\pi}\rp^{n_{12}} \lp - \frac{g^2 N l_1
l_3}{4\pi}\rp^{n_{13}} \lp \frac{g^2 N l_2 l_3}{4\pi}\rp^{n_{23}}\no\\
&& \times \frac1{(n_{12}+n_{13})!(n_{13}+n_{23})!(n_{12}+n_{23})!}
\lp \frac{n_{13}n_{23}}{|p_{1}|^2  |p_{2}|^2} +
\frac{n_{12}n_{23}}{|p_{1}|^2  |p_{3}|^2}+ \frac{n_{12}n_{13}}{|p_{2}|^2
|p_{3}|^2} \rp \,.
\eeeq   
A remarkable property of eq.~\re{fourier2} is that it displays a
strikingly simple dependence on the topology of the graph, namely on
the integers $n_{ij}$. 

In order to resum leading contributions in $N$, 
we just need to know the number of different configurations with fixed 
$n_{ij}$. We have already emphasized that cyclic permutations inside
any trace of eq.~\re{lead} do not alter the power of $N$, so that we
can conclude that the multiplicity of planar graphs amounts to
$(n_{12}+n_{13})(n_{13}+n_{23})(n_{12}+n_{23})$.
At this stage the full contribution to the correlator
$<W(p_1)W^{\dagger}(p_2)W(p_3)>$ can be readily written
\beeq
\label{3cor}
&&<W(p_1)W^{\dagger}(p_2)W(p_3)>=
\sum_{n_{12},n_{23},n_{13}=1} 
\frac{(-1)^{n_{13}}}{\pi^2 N} \, \de^{(2)}(p_1+p_2+p_3) \\
&& \times \frac{\ta_1^{n_{12}+n_{13}} \, \ta_2^{n_{12}+n_{23}} \,
\ta_3^{n_{13}+n_{23}}}{(n_{12}+n_{13}-1)!(n_{13}+n_{23}-1)!(n_{12}+n_{23}-1)!}
\lp \frac{n_{13}n_{23}}{|p_{1}|^2  |p_{2}|^2} +
\frac{n_{12}n_{23}}{|p_{1}|^2  |p_{3}|^2}+ \frac{n_{12}n_{13}}{|p_{2}|^2
|p_{3}|^2} \rp \,, \no
\eeeq
where $\ta_i=\sqrt \frac{g^2 N l_i^2}{4\pi}$.

Finally, we have to take into account the effect of  diagrams
with ``bubbles'', namely with propagators beginning and ending on the
same line. Again, only leading configurations with respect to $N$ are
to be considered. The counting of such configurations has already been performed
in \cite{noi}. The number of ways in which we can form $q$ pairs on a line
while keeping the leading power in $N$ turns out to be 
\beq
\label{bub} 
S_q=\frac{2^{2q}}{(q+1)!}\frac{\G(q+\half)}{\G(\half)} \,.
\eeq
On the other hand, the geometric factor associated with propagators starting and 
ending on the same line
factorizes and amounts to $\lp -N/(4\pi) \rp^{q}$ \cite{noi}.

Carefully inserting these factors, taking into account the new
multiplicity of planar  diagrams and then summing over the number of pairs,
we are led to replace eq.~\re{3cor} with the following 
three expressions, according to the number of lines which are affected.

In the case of bubbles on all the three lines we get
\beeq
\label{3bub}
&&<W(p_1)W^{\dagger}(p_2)W(p_3)>_{\PP \PP \PP}=
\sum_{n_{12},n_{23},n_{13}=1} 
\sum_{q_1,q_2,q_3=1} 
\frac{(-1)^{n_{13}}}{\pi^2 N} \, 
\de^{(2)}(p_1+p_2+p_3)  \\
&& \times 
\,(-1)^{q_1+q_2+q_3} 
\frac{\ta_1^{n_{12}+n_{13}+2 q_1} \, 
\ta_2^{n_{12}+n_{23}+2 q_2} \,  
\ta_3^{n_{13}+n_{23}+2q_3}}{(n_{12}+n_{13}+2 q_1-1)!(n_{13}+n_{23}+2
q_2-1)!(n_{12}+n_{23}+2 q_3-1)!} \, 
\no \\
&& \times 
\,S_{q_1} S_{q_2} S_{q_3}
(n_{12}+n_{13}) (n_{12}+n_{23}) (n_{13}+n_{23})
\lp \frac{n_{13}n_{23}}{|p_{1}|^2  |p_{2}|^2} +
\frac{n_{12}n_{23}}{|p_{1}|^2  |p_{3}|^2}+ \frac{n_{12}n_{13}}{|p_{2}|^2
|p_{3}|^2} \rp \no \,. 
\eeeq
In the case of bubbles on two lines
\beeq
\label{2bub}
&&<W(p_1)W^{\dagger}(p_2)W(p_3)>_{\PP \PP }=
\de^{(2)}(p_1+p_2+p_3) 
\sum_{n_{12},n_{23},n_{13}=1} 
\sum_{q,r=1} 
\frac{(-1)^{n_{13}}}{\pi^2 N} \, (-1)^{q+r} \, S_{q} S_{r} 
\no \\
&& \times \ta_1^{n_{12}+n_{13}}
\, \ta_2^{n_{12}+n_{23}}\,\ta_3^{n_{13}+n_{23}}
\lq  \frac{\ta_1^{2 q}
\, \ta_2^{2r}}{(n_{12}+n_{13}+2q-1)!
(n_{12}+n_{23}+2r-1)!(n_{13}+n_{23})!}
\right. \no \\  
&&\ph{\times \ta_1^{n_{12}+n_{13}}
\, \ta_2^{n_{12}+n_{23}}\,\ta_3^{n_{13}+n_{23}}}
+\frac{\ta_1^{2 q} \,\ta_3^{2r}}{(n_{12}+n_{13}+2 q-1)!
(n_{12}+n_{23})! (n_{13}+n_{23}+2r-1)!}  \no \\
&&\ph{\times \ta_1^{n_{12}+n_{13}}
\, \ta_2^{n_{12}+n_{23}}\,\ta_3^{n_{13}+n_{23}}}
\left.+
\frac{\ta_2^{2 q} \, 
\ta_3^{2r}}{(n_{12}+n_{13})!(n_{12}+n_{23}+2q-1)!
(n_{13}+n_{23}+2r-1)!}\rq \no \\
&& \times
(n_{12}+n_{13}) (n_{12}+n_{23}) (n_{13}+n_{23})  
\lp \frac{n_{13}n_{23}}{|p_{1}|^2  |p_{2}|^2} +
\frac{n_{12}n_{23}}{|p_{1}|^2  |p_{3}|^2}+ \frac{n_{12}n_{13}}{|p_{2}|^2
|p_{3}|^2} \rp   \,, 
\eeeq
and finally, when only one line is affected,
\beeq
\label{1bub}
&&<W(p_1)W^{\dagger}(p_2)W(p_3)>_{\PP}=
\de^{(2)}(p_1+p_2+p_3) 
\sum_{n_{12},n_{23},n_{13}=1} 
\sum_{q=1} 
\frac{(-1)^{n_{13}}}{\pi^2 N} \, (-1)^{q} \, S_{q}  
\no \\
&& \times \ta_1^{n_{12}+n_{13}}
\, \ta_2^{n_{12}+n_{23}}\,\ta_3^{n_{13}+n_{23}}
\lq  \frac{\ta_1^{2 q}}{(n_{12}+n_{13}+2q-1)!
(n_{12}+n_{23})!(n_{13}+n_{23})!}
\right. \no \\  
&&\ph{\times \ta_1^{n_{12}+n_{13}}
\, \ta_2^{n_{12}+n_{23}}\,\ta_3^{n_{13}+n_{23}}}
+\frac{\ta_2^{2 q}}{(n_{12}+n_{13})!
(n_{12}+n_{23}+2q-1)! (n_{13}+n_{23})!}  \no \\
&&\ph{\times \ta_1^{n_{12}+n_{13}}
\, \ta_2^{n_{12}+n_{23}}\,\ta_3^{n_{13}+n_{23}}}
\left.+
\frac{\ta_3^{2 q}}{(n_{12}+n_{13})!(n_{12}+n_{23})!
(n_{13}+n_{23}+2q-1)!}\rq \no \\
&& \times
(n_{12}+n_{13}) (n_{12}+n_{23}) (n_{13}+n_{23})  
\lp  \frac{n_{13}n_{23}}{|p_{1}|^2  |p_{2}|^2} +
\frac{n_{12}n_{23}}{|p_{1}|^2  |p_{3}|^2}+ \frac{n_{12}n_{13}}{|p_{2}|^2
|p_{3}|^2} \rp   \,.
\eeeq

\section{Asymptotic behaviour}

Since we are interested in large values of the variables $\ta_i$, the series
above have to be explicitly resummed. 
This is most easily done by recalling the inverse Laplace
transform 

$$\frac {\ta ^n}{n!}=\int_{c-i\infty}^{c+i\infty} \frac {dw}{2\pi i} 
w^{-n-1} e^{w\ta},\qquad\qquad c>0.$$

The typical sums in eq.~\re{3cor} are
\beeq \label{typ}
&&{\cal S}\equiv \sum_{n_{12},n_{23},n_{13}=1} (-1)^{n_{13}} 
\frac{\ta_1^{n_{12}+n_{13}-1} \, \ta_2^{n_{12}+n_{23}-1} \,
\ta_3^{n_{13}+n_{23}-1}}{(n_{12}+n_{13}-1)!(n_{13}+n_{23}-1)!(n_{12}+n_{23}-1)!}\\ \no
&&=-\int_{c-i\infty}^{c+i\infty}\frac {dw_1 dw_2 dw_3}{(2\pi i)^3}
\frac{e^{w_1\ta_1+w_2\ta_2+w_3\ta_3}}{(w_1w_2-1)(w_2w_3-1)(w_3w_1+1)}\,,\qquad
c>1.
\eeeq

A careful treatment of those integrals leads to the asymptotic estimate
\beq \label{asy}
{\cal S}\simeq -\exp(2\sqrt{\ta_2(\ta_1+\ta_3)}),
\eeq
apart from slowly increasing power factors. A detailed calculation 
is reported in appendix B.

Then the extra $n_{ij}$ factors in eq.~\re{3cor} can be accounted for by suitable
differentiations; 
for instance the factor $n_{12}$ can be obtained
taking derivatives with respect to the variable $\xi_{12}\equiv \ta_1
\ta_2$. 
A little thought is enough to conclude
that the leading term occurs when the derivatives act on the exponential
factor, so that one can differentiate directly the asymptotic estimate
and only the power factor in eq.~\re{asy} is changed.
Actually, owing to the alternating sign with $n_{13}$ in eq.~\re{3cor}, 
the other two terms, namely the ones involving $n_{13}$, turn out to
be subleading. 

Taking the relation eq.~\re{leng} into account and remembering that the
two-line correlator 
$<W(p)W^{\dagger}(p)>$ increases like $\exp(2\ta)$, 
one concludes that
the {\it normalized} three-line correlator goes to a constant at $\infty$
(always disregarding  power corrections).

Two remarkable features are to be noticed. 

First, the increase 
of the correlator with  increasing lengths of the lines,
in spite of the fact that an interference occurs in 
the exchanges between $W(p_1)$ and $W(p_3)$. This effect is overwhelmed
by the coherent exchanges between $W(p_1)$ and $W^{\dagger}(p_2)$, 
$W^{\dagger}(p_2)$ and $W(p_3)$. The increase is therefore a consequence
of the parallelism of the lines, as correctly suggested in \cite{gross}.

The second effect, which has not been noticed in previous treatments,
is the extra $1/N$ factor; in the case of a $\nu$-line correlator
this factor would be $N^{2-\nu}$. As a consequence, multiple line
correlators get more and more depressed in the 't Hooft limit.

In the case of lines with bubbles (see {\em e.g.} eq.~\re{3bub}), the typical sums in eq.~\re{typ} are thereby to be replaced with
\beeq \label{typp}
&&{\cal S}\equiv \sum_{n_{12},n_{23},n_{13}=1} \sum_{q_1,q_2,q_3=1} \frac{(-1)^{n_{13}} 
(-4)^{q_1+q_2+q_3}}{\pi \sqrt {\pi}} \frac{\G(q_1+1/2)}{(q_1+1)!}
\frac{\G(q_2+1/2)}{(q_2+1)!} \frac{\G(q_3+1/2)}{(q_3+1)!}\no \\
&&\times \frac{\ta_1^{n_{12}+n_{13}+2q_1-1} \, \ta_2^{n_{12}+n_{23}+2q_2-1} \,
\ta_3^{n_{13}+n_{23}+2q_3-1}}{(n_{12}+n_{13}+2q_1-1)!(n_{13}+n_{23}+2q_2-1)!
(n_{12}+n_{23}+2q_3-1)!}\, .
\eeeq

Introducing again inverse Laplace transforms we get
\beeq \label{tylap}
&&{\cal S}= \sum_{n_{12},n_{23},n_{13}=1} \sum_{q_1,q_2,q_3=1} \frac{(-1)^{n_{13}} 
(-4)^{q_1+q_2+q_3}}{\pi \sqrt {\pi}} \frac{\G(q_1+1/2)}{(q_1+1)!}
\frac{\G(q_2+1/2)}{(q_2+1)!} \frac{\G(q_3+1/2)}{(q_3+1)!}\no \\
&&\times \int_{c-i\infty}^{c+i\infty} dw_1 dw_2 dw_3 e^{w_1\ta_1+w_2\ta_2+w_3
\ta_3} w_1^{-(n_{12}+n_{13}+2q_1)}  w_2^{-(n_{23}+n_{12}+2q_2)}
w_3^{-(n_{13}+n_{23}+2q_3)}\no \\
&&=- \sum_{q_1,q_2,q_3=1} \frac{ 
(-4)^{q_1+q_2+q_3}}{\pi \sqrt {\pi}} \frac{\G(q_1+1/2)}{(q_1+1)!}
\frac{\G(q_2+1/2)}{(q_2+1)!} \frac{\G(q_3+1/2)}{(q_3+1)!}\no \\
&&\times \int_{c-i\infty}^{c+i\infty}\frac {dw_1 dw_2 dw_3}{(2\pi i)^3}
\frac{e^{w_1\ta_1+w_2\ta_2+w_3\ta_3}}{(w_1w_2-1)(w_2w_3-1)(w_3w_1+1)}
w_1^{-2q_1}w_2^{-2q_2}w_3^{-2q_3}\, ,
\eeeq
with $c>1.$

The extra sums
\beq \label{hyper}
\sum_{q=1}\frac{\G(q+1/2)}{\G(q+2)}\frac{(-4w^{-2})^q}{\sqrt {\pi}}=
\frac{2w}{w+\sqrt{w^2+4}}-1 \equiv \phi (w)
\eeq
do not produce any singularity for Re$\, w>0.$
As a consequence they are ineffective on the asymptotic behaviour at large $\ta_i$.

Coming back to eqs.~\re{3bub}, \re{2bub} and \re{1bub}, additional 
factors of
the kind $n_{ij}n_{kl}$ can be accounted for by suitable differentiations,
while factors like $n_{ij}+n_{kl}$ can be cancelled, for instance, by means of a shift in
the denominators, {\it viz}
\beq \label{shift}
\frac{n_{12}+n_{13}}{(n_{12}+n_{13}+2 q_1-1)!}=\frac{1}{(n_{12}+n_{13}+2 q_1-2)!}-
\frac{2q_1-1}{(n_{12}+n_{13}+2 q_1-1)!}\, .
\eeq
Both procedures entail the appearance of extra powers of $q_i$ in the numerator,
which however are ineffective on the analytic behaviour of eq.~\re{hyper}, owing to the
alternating sign of the sums over $q_i$.
Considering for instance the equation above,
the net result is an increase of powers in $\ta_i$ 
coming from the former addendum, whereas
the latter can be disregarded, remaining subleading.

After a careful examination of all the terms coming from such a proliferation,
it turns out (see appendix B) that the largest contribution at large $N$ and $\ta_i$
comes from diagrams with bubbles on all the three lines and
$n_{13}=0.$ It increases  (in absolute value) like
\beq\label{fin}
<W(p_1)W^{\dagger}(p_2)W(p_3)>\simeq - \frac{\delta^2(p_1+p_2+p_3)}{|p_1|^2|p_3|^2} 
\ta^{15/2} \exp{(2\ta)},
\eeq
where we have set $\ta\equiv \ta_2=\ta_1 +\ta_3$, according to eq.~\re{leng}, and
chosen, for the sake of simplicity, $\ta_1=\ta_3$.

If we now remember that the two-line correlator increases with $\ta$
like \cite{noi}~\footnote{The infinite normalization $\de (0)$ coming
from momentum conservation will here be dropped.}
\beq \label{twol}
<W(p)W^{\dagger}(p)> \simeq \ta^{7/2} \exp {(2\ta)},
\eeq
we can easily realize that the {\it normalized} three-line correlator, in the case
of parallel lines, increases (in absolute value) mildly with $\ta$, 
namely as $\ta^{9/4}$, for fixed values of the momenta.

We notice that, although eqs.~\re{fin} and \re{twol} followed from a
perturbative analysis, having resummed all orders, they hold also at
large $g^2N$ ($g$ fixed). This fact protects the normalized three-line 
correlator against suppression at large $N$.

Unfortunately, if the lines are not parallel, the considerations concerning
 colour structure can be repeated, but the result eq.~\re{vvv} exhibits
an explicit dependence on the variables $a_i,b_j$ and $c_k$. In addition
such a dependence 
is different for different graphs and this eventually prevents us from
performing  integrations over the line variables at a generic
perturbative order (the analog of eq.~\re{fourier2}). 

Owing to this limitation, we are unable to draw any concrete prediction
concerning non-parallel lines, while expecting, on a purely intuitive basis,
a loss of coherence and thereby a decrease with the line lengths.

\section{Conclusions}

In this paper a perturbative calculation of the correlator of three
parallel open Wilson lines
in two non-commutative space-time dimensions was performed. In the
large-$N$ planar limit, the perturbative series was fully resummed.
Remarkably, $\th$-dependence turns out to be trivial, intervening only
through the length of the lines, just in those diagrams which are
dominant at large $N$. As pointed out in \cite{noi}, this
feature is peculiar of two dimensions, where transverse degrees of
freedom are absent and the theory exhibits invariance under
area-preserving diffeomorphisms, and in fact the same result was
previously found for the correlator of two
parallel Wilson lines.

Being interested in the large argument behaviour of the three-line
correlator, we were able to give an asymptotic estimate, showing
an exponential growth of the correlator with the
lengths of the lines, in spite of an  interference effect between
lines with the same orientation. This result generalizes a similar
increase occurring in the two-line correlator, so that we could
conclude that the {\it normalized} three-line correlator is still
increasing like a (small) power of its argument.

A novel feature of our treatment is the appearance of a damping factor 
$1/N$, which  implies that the correlator considered is depressed in
the 't Hooft limit, at odds with the two-line analog. As the number of 
lines involved is increased, multiple line correlator are expected to
get more and more depressed.

Actually, our treatment can, in principle, be extended to multiple (parallel) line correlators.
The first novelty concerns the colour structure. A simple thought is enough to
conclude that the colour pattern of eq.~\re{lead} is to be generalized
to {\it cyclic colour strings}, namely strings with lines
exchanging propagators only between nearest neighbours. As an example with
$\nu$ lines, we consider the pattern
\beeq \label{cyclead}
&&(a_1\ldots a_{n_{\nu 1}}b_1\ldots b_{n_{12}})(b_{n_{12}}\ldots b_1 c_1\ldots
c_{n_{23}})(c_{n_{23}}\ldots c_1 d_1\ldots d_{n_{34}})\no \\
&&\ldots (z_1\ldots 
z_{n_{\nu-1\nu}}a_{n_{\nu 1}}\ldots a_1)=N^{2-\nu}(N/2)^n, 
\eeeq
$n$ representing, as usual, the total number of  propagators.
Notice the power $N^{2-\nu}$, which shows how {\it connected} configurations
are subleading with respect to disconnected ones. Indeed, the latter
can be realized as a product of the former. For the sake of clarity,
for the product of two connected correlators with $\n_1$, $\n_2$ lines,
respectively, at the perturbative order $n=n_1+n_2$, from
eq.~\re{cyclead} we can infer the behaviour
\beq \label{prod2}
N^{2-\nu_{1}}(N/2)^{n_1} N^{2-\nu_{2}}(N/2)^{n_2}=
N^{4-\nu}(N/2)^{n}\,, \qquad \quad  \n=\n_1+\n_2\,.
\eeq
Cyclic configurations as in eq.~\re{cyclead}
are colour-leading, but  are not the only ones. For instance, with $\nu=4$,
the configuration
\beeq\label{lead4}
&&(a_1\ldots a_{n{41}}f_1\ldots f_{n_{13}}b_1\ldots b_{n_{12}})
(b_{n_{12}}\ldots b_1 c_1\ldots c_{n_{23}})\no \\
&&\times (c_{n_{23}}\ldots c_1f_{n_{13}}\ldots f_1 d_1\ldots d_{n_{34}})
(d_{n_{34}}\ldots d_1 a_{n_{41}}\ldots a_1)= N^{-2}(N/2)^n\,,
\eeeq
is still leading, whereas
\beeq\label{sub4}
&&(a_1\ldots a_{n{41}}f_1\ldots f_{n_{13}}b_1\ldots b_{n_{12}})
(b_{n_{12}}\ldots b_1g_1\ldots g_{n_{24}} c_1\ldots c_{n_{23}})\\
&&\times (c_{n_{23}}\ldots c_1f_{n_{13}}\ldots f_1 d_1\ldots d_{n_{34}})
(d_{n_{34}}\ldots d_1 g_{n_{24}}\ldots g_1a_{n_{41}}\ldots a_1)= N^{-4}(N/2)^n
\no 
\eeeq
is not.
For $\nu=4$ the colour leading configurations can be counted; then the 
``geometric'' structure can be worked out, suitably generalizing the
treatment leading to eq.~\re{generic}. After a long (and cumbersome)
calculation, indications emerge of a coherent increase with the line lengths
of the kind already described.

Beyond $\nu=4$, both the singling out of the colour leading configurations
and the subsequent evaluation of the geometrical factors become almost
intractable. On the other hand, in our opinion, such an effort would not
be worthwhile since it could hardly add  new understanding
from the point of view of the physics involved.  

A different consideration might be deserved by the possibility of performing
non-perturbative evaluations based on a compaction 
of the theory
on a torus, followed by a Morita mapping, repeating the treatment
developed in \cite{noi} for the two-line correlator. Here the difficulty
concerns the harmonic analysis on the commutative torus when three (or more)
cycles are involved and the subsequent identification of suitable
saddle points (if any), in order to concretely evaluate the large $N$,
large $\ta_i$ limit. It would be nice to discover a  coherent
increase of multiple line correlators in a non-perturbative context.
 
These and related developments will be deferred to future investigations
and the results will be reported elsewhere.

\acknowledgments
Discussions with A. Torrielli are gratefully acknowledged. One
of us (F.V.) wishes to thank  the CERN-TH Division for hospitality
during the completion of this work.

\appendix

\section{Details of calculations appearing in section 3}

Starting from eq.~\re{generic}, we notice that the polynomial ${\cal P}$ in the 
Moyal phase 
$e^{i{\cal{P}}(P_i,Q_j,R_k; \theta,{\bf p})}$  
is (for the relevant class of graphs) at most
linear in $P_i,Q_j,R_k$ and, since $p_r\theta p_s\equiv 0$,
 the constant term is missing. We can then factorize
$e^{i\cal P}$, provided we replace 
$\xi_i -{\zeta}_i,{\eta}_j-{\xi}'_j,\eta'_k-\zeta'_k$ with  
$a_i,b_j,c_k$ defined as in eq.~\re{fourier}.
Furthermore, we introduce three complex variables $U,V,W$ in order to 
express the $\delta$-functions enforcing momentum conservation as 
 integrals, {\em e.g.}
\beq
\label{coccoa}
\de^{(2)}(\sum_i P_i +\sum_k R_k -p_1) \, =
(2\pi)^{-2}\int{d^2 U e^{i(\sum_i P_i +\sum_k R_k -p_1)\cdot U}}\,.
\eeq
We end up with an expression of the form
\beq
(2\pi)^{-6}\int{e^{-i(p_1\cdot U +p_2 \cdot V + 
p_3 \cdot W)}f(U-V,V-W,W-U)d^2U d^2V d^2W}\, ;
\eeq
it can be rewritten as
\beeq
&&(2\pi)^{-6}\int 
e^{-\frac{i}{3}(p_1+p_2+ p_3)
\cdot (U +V +W)}[e^{-\frac{i}{3}((p_1-p_2)\cdot (U-V)+(p_2-p_3)\cdot (V-W)+
(p_3-p_1)\cdot(W-U))}\nonumber\\
&& \ph{(2\pi)^{-6}
}
\times f(U-V,V-W,W-U)]\,d^2U d^2V d^2W\,,
\eeeq
from which the expected  momentum conserving $\delta$-function factorizes
\beeq
&&(2\pi)^{-4}\delta^{(2)} (p_1+p_2+p_3)\int 
[e^{-\frac{i}{3}((p_1-p_2)\cdot x+
(p_2-p_3)\cdot y+
(p_3-p_1)\cdot z)}f(x,y,z)]\no \\ 
&& \ph{(2\pi)^{-4}\delta^{(2)} (p_1+p_2+p_3)
}\times \delta^{(2)}(x+y+z)\,d^2x \,d^2y \,d^2z\,.
\eeeq
Now we introduce a two component variable $P$ to exponentiate 
$\delta^{(2)}(x+y+z)$,
as in eq.~\re{coccoa}, and then perform the integrations over
$P_i,Q_j,R_k$. By recalling eq.~\re{euprop}, 
we finally obtain the factorized form eq.~\re{fourier}. 

Next, we have to deal with three integrals of the form
\beq
\label{coccoe}
I(P,\tilde{p}_1,{\bf a})\equiv \int{d^2 x \, e^{i(P-\tilde{p}_1)\cdot x}
\prod_{i=1}^{n_{12}} \frac{x-a_i}{(x-a_i)^*}}
\eeq
(see eq.~\re{intp}).
It is understood here that we must perform a symmetric integration around the 
poles $x=a_i$; hence we need to decompose the integrand in eq.~\re{coccoe} in 
simple fractions. We can use almost verbatim a result from 
\cite{noi}. The 
denominator decomposes as in eq.~(8.2) therein, apart from minor variable 
redefinitions
\beq
\prod_{i=1}^{n_{12}} \frac{1}{(x-a_i)^*}=
\sum_{i=1}^{n_{12}} \frac{1}{(x-a_i)^*} 
\prod_{j\neq i}\frac{1}{(a_i-a_j)^*} \,.
\eeq
In each term we shift the integration variable as  
$x-a_i\to x$, expand the  polynomial in the numerator in the new variable $x$, 
perform a symmetric integration according to
\beq
\label{distributionlike}
\int d^2x \,e^{-i P\cdot x}\,\frac{x^m}{x_-}=\frac{\pi (-2i)^{m+1}m!}{P_-^{m+1}}
\eeq
and obtain eq.~\re{intp}\footnote{The integrals in eqs.~\re{coccoe} and \re{distributionlike} 
are to be 
intended in the sense of the theory of distributions. }.

Let us now consider the integral 
\beq
\label{vvvv}
\int d^2P\, \frac{e^{iP\cdot (a_i+b_j+c_k)}}{ (P-\tilde{p}_{1})_-^{m_1+1}
(P-\tilde{p}_{2})_-^{m_2+1} (P-\tilde{p}_{3})_-^{m_3+1}}\,,
\eeq
which has to be computed via symmetric integration around the singularities 
of the integrand. 
Again, we need to decompose the integrand in simple fractions, which can 
subsequently be integrated by means of standard techniques. 
Hence we must find suitable coefficients $c_{r,n_r}$ so as to match
\beq
\label{coccoint}
\frac{1}{ (P-\tilde{p}_{1})^{m_1+1}_-
(P-\tilde{p}_{2})^{m_2+1}_- (P-\tilde{p}_{3})^{m_3+1}_-} = \sum_{r=1}^{3} 
\sum_{n_r =1}^{m_r +1} c_{r,n_r} 
\frac{1}{ (P-\tilde{p}_{r})^{n_r}_-} \,.
\eeq 
The coefficients $c_{r,n_r}$ can be obtained by imposing that both sides of
eq.~\re{coccoint}, when multiplied by an arbitrary power of 
$(P-\tilde{p}_{i})$, have coincident residues at each pole $P=\tilde{p}_{i}$.
The result is
\beq
\label{coccoc}
c_{r_,n_r}=\sum_{\{l_s\,:\, l=m_r-n_r\}}
(-1)^l
\prod_{s\neq r} 
{m_s + l_s \choose l_s}
\frac{1}{(\tilde{p}_r - \tilde{p}_s)^{m_s + l_s +1}_-}\,,
\eeq
where $l\equiv\sum_{s\neq r}l_s$\,.
The single terms in the decomposition of eq.~\re{coccoint} can then be 
symmetrically
integrated around their pole, the result being 
\beq
\int d^2 P \frac{e^{iP\cdot (a_i+b_j+c_k)}}{ (P-\tilde{p}_{r})^{n_r}_-}=
e^{i\tilde{p}_r\cdot(a_i+b_j+c_k)} \frac{\pi i^{n_r} 2^{-n_r +2}}{(n_r -1)!}
\frac{(a_i+b_j+c_k)^{n_r -1}}{(a_i+b_j+c_k)^{*}}\,,
\eeq
whence eq.~\re{vvv} follows.

We can now single out  in eq.~\re{vvv} the only relevant terms.
Since all the  
parameters have the same complex phase (owing to the parallelism of the 
three Wilson lines), it is easily seen that the phase of each summand in 
eq.~\re{vvv} depends 
only on $n_{12},n_{13},n_{23}$, which are fixed for a given graph, but not on 
any of the indices which are summed over. 
After factorizing a common phase~\footnote{Such a phase will
eventually cancel against an opposite one originating from the
integration over the line variables \cite{noi}.}, we can replace
$a_i^*$, $b_j^*$, $c_k^*$ with $a_i$, $b_j$, $c_k$, respectively, and
$\tilde{p}_i$ with $|p_i|$  in eqs.~\re{v} 
and \re{vvv}.
Hence, on one hand eq.~\re{vvv} is symmetric under the exchange of 
any pair of $a_i$, or $b_j$, or $c_k$, on the other hand, the 
common denominator in the sum over $i,j,k$, being the product of the
three Vandermonde determinants built with $a_i,b_j,c_k$, is antisymmetric. 
It follows that
the numerator of the sum must be antisymmetric too, and this entails 
(by repeated use of Ruffini's theorem) that 
it has those determinants as factors and therefore the sum must have non-negative 
degree in the parameters. Since every term has manifestly non-positive 
degree, it follows that only zero degree terms survive.
One can realize that those originate \eg from the first term  of the r.h.s. of
eq.~\re{p-int}   when the conditions $m_r=1$, $l_r=0$, $r=2,3$ are fulfilled,
and  from the other two terms when analogous conditions hold. 

They can be evaluated as follows. Let us consider for the sake of
example again the first contribution to the function $\KK$ appearing
in eq.~\re{p-int}. When inserted in eq.~\re{vvv}, 
two out of three 
functions $V$ 
are forced to equal unity, namely those depending on
${\bf b}$ and ${\bf c}$, since $m_2=m_3=1$.
The relevant sum ({\em i.e.} apart from factors that are independent of the 
indices summed over) reduces to
\beq
\label{coccod}
\sum_{m_1=1}^{n_{12}}(-1)^{m_1} \sum_{i=1}^{n_{12}}\sum_{j=1}^{n_{13}}
\sum_{k=1}^{n_{23}}
\sum_{j_1<j_2<\ldots<j_{n_{12}-m_1}}
\frac{(a_i - a_{j_1})\ldots (a_i - a_{j_{n_{12}-m_1}})}{\prod_{j\neq
i}(a_i-a_j)} 
(a_i +b_j+c_k)^{m_1 -1}\,.
\eeq
When we sum over $i$ while keeping $j,k$ and $m_1$ fixed, the result must be 
symmetric under the exchange of any pair of $a_i$; hence (by the same 
token as above) only zero 
degree terms survive, and we can set $b_j$, $c_k$ equal to zero, as 
they multiply negative powers of $a_i$. The sum over $n_{13}$ and
$n_{23}$ then merely results in a multiplicity factor $n_{13}n_{23}$.

Now in the equivalent expression
\beq
\sum_{i=1}^{n_{12}} \sum_{j_1<j_2<\ldots<j_{m_1 -1}}
\frac{a_i^{m_1 -1}}{(a_i-a_{j_1})\ldots (a_i-a_{j_{m_1 -1}})}
\eeq
we can see that each term depends on exactly $m_1$ out of the $n_{12}$ 
parameters $a_i$.

There are just ${n_{12} \choose m_1}$ 
such sets, and to each of them again the same argument applies, allowing 
only zero degree terms to survive. These are pure numbers, and can actually be 
seen to equal unity (by means of counting  the occurrences of a given 
monomial in the numerator and in the denominator). Finally, eq.~\re{coccod} becomes
\beq
\sum_{m_1=1}^{n_{12}}\left(\begin{array}{c}n_{12}\\m_1\end{array}\right)(-1)^{m_1}n_{13}n_{23}=-n_{13}n_{23}\,.
\eeq
Analogous results are found when the other two terms contributing to the
function $\KK$ in eq.~\re{p-int} are considered, so that we can
conclude eq.~\re{fourier2} follows.

\section{Asymptotic estimates}

We can infer the asymptotic behaviour of the various contributions eqs.~\re{3cor},
\re{3bub} through \re{1bub} for large $\tau_i$ from the 
Laplace transform representation.
In what follows, it is understood that the variables $\tau_i$ are sent to infinity, 
while keeping  their ratios finite ({\em e.g. } $\tau_1=\tau_3=\frac{\tau_2}{2}$).
We proceed to estimate the inverse Laplace transform by looking at each 
stage for the rightmost singularity, which will give the exponentially 
dominant contribution.
Since we are not considering disconnected graphs, we have to deal separately 
with contributions where  $n_{12},n_{13},n_{23}\neq 0$, and with those 
where one (and one only) among $n_{12},n_{13},n_{23}$ vanishes. Besides, for 
each case, there are contributions arising from graphs without self energy 
bubbles, and with bubbles on one, two, or all the three Wilson lines. Here 
we will sketch how 
the asymptotic estimate is achieved in the simplest cases (no bubbles), and 
in the dominant case, which turns out to consist of graphs with bubbles on 
three lines, and either $n_{12},n_{13},n_{23}\neq 0$ or only $n_{13}=0$.

Let us start from the case without bubbles. In eq.~\re{typ} we can integrate first 
over $w_1$, the sign of the exponential allowing us to shift the contour to 
the left; the singularities are in $w_1=\frac{1}{w_2}$ and 
$w_1=-\frac{1}{w_3}$. 
Since $w_2,w_3$ (as well as $w_1$) vary  along a vertical line with a 
real part $c>1$, the latter pole has negative real part and 
produces a negligible (exponentially suppressed) contribution when 
compared to the former; we will henceforth drop this and similar terms 
altogether.
Hence
\beq
{\cal{S}} \simeq  -\int_{c-i\infty}^{c+i\infty}
\frac {dw_2 dw_3}{(2\pi i)^2}
\frac{e^{\frac{\ta_1}{w_2} +w_2\ta_2+w_3\ta_3}}{(w_2 + w_3)(w_3 w_2+1)}\,,
\qquad c>1\,.
\eeq 
Next we integrate over $w_3$, and drop the pole in $w_3=-w_2$
\beq
\label{coccosad}
{\cal{S}} \simeq - \int_{c-i\infty}^{c+i\infty}
\frac {dw_2 }{(2\pi i)}
\frac{e^{\frac{(\ta_1 +\ta_3)}{w_2} +w_2\ta_2}}{(w_2^2 +1)}\,,
\qquad c>1\,.
\eeq
The integral in eq.~\re{coccosad} can be approximated through the saddle point 
method: 
one finds the stationary point of the quadratic form in the exponent, which 
turns out to be 
$w_2=
\frac{\ta_1 +\ta_3}{\ta_2}$, then translates the contour to
$c=\frac{\ta_1 +\ta_3}{\ta_2}$, and  
wisely expands the integrand about the stationary point (that is, both the 
factor multiplying the exponential and the {\em exponent} are Taylor 
expanded, the quadratic part in the exponent gives rise to a Gaussian 
integration, higher terms are Taylor expanded if needed). In this 
way we get
\beq
{\cal S} \simeq -i \sqrt{\pi}\frac{(\ta_2(\ta_1+\ta_3))^{\frac{1}{4}}}{\ta_2}
e^{2(\ta_2(\ta_1+\ta_3))^{\frac{1}{2}}}\left(\frac{1}{\frac{\ta_1+\ta_3}{\ta_2}+1} +\ldots\right)\,,
\eeq 
where the neglected terms are suppressed by at least a factor of $1/\ta_i$.
It is very useful to notice the following point: the asymptotics of
eq.~\re{3cor}  
contains {\em derivatives} of ${\cal S}$ with respect to $\tau_i$, 
evaluated on the constraint $\tau_1 +\tau_3\equiv \tau_2$.
Indeed the  $n_{12}n_{13}$, $\ldots$  factors can be represented 
introducing new variables $\xi_{ij}\equiv \ta_i \ta_j\,,i<j$, as 
$n_{ij}\to \xi_{ij}\partial_{\xi_{ij}}$.

We should first differentiate and then impose the constraint. Nevertheless, in 
the spirit of this approximation, it is easy to realize that when  
any number of derivatives act on anything but 
$e^{2(\tau_2(\tau_1+ \tau_3))^{\frac{1}{2}}}$ in ${\cal S}$, 
they produce terms which are power suppressed in $\tau_i$.
Hence we are free to 
impose the constraint on everything but the exponential, and differentiate the 
exponential alone when evaluating the asymptotics of eq.~\re{3cor}. With this convention in mind, 
we can write as well
\beq
{\cal S} \simeq -i \frac{\sqrt{\pi}}{2\sqrt{\ta_2}}
e^{2(\ta_2(\ta_1+\ta_3))^{\frac{1}{2}}} +\ldots \,.
\eeq
Then, keeping in mind that derivatives are relevant when acting on the 
exponential, we see that only the term $\frac{n_{12}n_{23}}{p_{1-}^2 p_{3-}^2}$
contributes in the asymptotic limit. 

When we consider graphs with $n_{13}\equiv 0$ and no 
bubbles, it is easy to see that replacing the sum over $n_{13}$ with the 
single $n_{13}=0$ term results in the absence of the 
factor $\frac{-1}{w_1w_3+1}$ in the Laplace transform. This factor
does not contribute  dominant 
poles in the Laplace inversion and only amounts to a constant when 
evaluated on the relevant singularities. Then in eq.~\re{coccosad} the
factor $\frac{-1}{w_2^2 +1}$  is replaced with $\frac{1}{w_2^2}$, and, since 
eventually the constraint $w_2=\frac{\ta_1+\ta_3}{\ta_2}=1$ is imposed, 
the term with $n_{13}=0$ is (in the large $\ta$ limit) minus twice
the term with $n_{12},n_{13},n_{23}\neq 0$. Terms with 
$n_{12}=0$ or $n_{23}=0$ are exponentially suppressed, since one of the 
leading singularities is missing.

When we come to contributions with bubbles on one or more lines, the 
asymptotic estimate is very similar, though more involved.
We consider the example with  $n_{12},n_{13},n_{23}\neq 0$ and bubbles on 
all the lines, all the other cases behaving analogously. 

First, when comparing eq.~\re{3bub} with eq.~\re {3cor}, we notice the
following differences:
\begin{itemize}
\item 
denominators like $\frac{1}{(n_{12}+n_{13}-1)!}$, $\ldots$ are replaced with 
$\frac{n_{12}+n_{13}}{(n_{12}+n_{13}+2q_1 -1)!}$, $\ldots$
\item 
the power of $\ta_i$ is increased by $2q_i$
\item
the factor  $(-1)^{(q_1+q_2+q_3)}S_{q_1}S_{q_2}S_{q_3}$  appears. 
\end{itemize}
Due to the modified powers of $\ta_i$, factors like $n_{12}$, $\ldots$ cannot be exactly 
reproduced by derivatives with respect to $\ta_i$, but
the mismatch is subleading. Indeed, replacing \eg $n_{12}$ with 
$\xi_{12}\partial_{\xi_{12}}$  means to approximate 
$n_{12}\simeq n_{12}+(q_1+q_2-q_3)$, and the difference is subleading 
because, as we shall see, $q_i$ factors can only affect the constant 
multiplying the leading exponential behaviour, while the derivatives 
$\xi_{ij}\partial_{\xi_{ij}}$ introduce powers of $\xi_{ij}$.
With this in mind, we can recover the various factors  $n_{ij}$ by 
differentiation, factor out $\ta_1 \ta_2 \ta_3$ as we did for eq.~\re{3cor}, 
and then Laplace transform. 
The sums over $q_i$ as in eq.~\re{hyper} 
have no pole. Multiplying $S_{q_i}$ by $q_i$ amounts to act on 
$\phi(w)$ with 
$-\frac{1}{2} w\frac{d}{dw}$, and again no poles are produced 
(hence justifying our former claim). The presence of the factors $\phi(w_i)$  
boils down, in the saddle point approximation, to powers of
$\phi(1)=\frac{\sqrt{5}-3}{2}$, while the extra derivatives contribute 
powers of $\ta_i$. The terms with more derivatives, {\em i.e.} more factors of 
$n_{ij}$, dominate; they come from bubbles on the three lines, 
both with $n_{ij}\neq0$ and with $n_{13}=0$, the latter being, by the same 
argument discussed above, twice the former and opposite in sign.

\end{document}